# Structural phase stability and Magnetism in $Co_2FeO_4$ spinel oxide

I. Panneer Muthuselvam and R.N. Bhowmik[*]

Department of Physics, Pondicherry University, R. Venkataraman Nagar, Kalapet, Pondicherry-605014, India

[*]E-mail address for corresponding author: rnbhowmik.phy@pondiuni.edu.in

Abstract

We report a correlation between structural phase stability and magnetic properties of $Co_2FeO_4$ spinel oxide. We employed mechanical alloying and subsequent annealing to obtain the desired samples. The particle size of the samples changes from 25 nm to 45 nm. The structural phase separation of samples, except sample annealed at $900^0C$, into Co rich and Fe rich spinel phase has been examined from XRD spectrum, SEM picture, along with EDAX spectrum, and magnetic measurements. The present study indicated the ferrimagnetic character of $Co_2FeO_4$, irrespective of structural phase stability. The observation of mixed ferrimagnetic phases, associated with two Curie temperatures at $T_{C1}$ and $T_{C2}$ ($>T_{C1}$), respectively, provides the additional support of the splitting of single cubic spinel phase in $Co_2FeO_4$ spinel oxide.

## I. INTRODUCTION

Spinel ferrites are represented by the formula unit $AB_2O_4$. Most of the spinel ferrites form cubic spinel structure with oxygen anions in fcc positions and cations in the tetrahedral and octahedral coordinated interstitial lattice sites, forming the A and B sublattices [1]. Depending upon the nature (magnetic or non-magnetic) and distribution of cations among A and B sublattices, spinel ferrites can exhibit properties of different type magnets, like: ferrimagnet, antiferromagnet and paramagnet. The inter-sublattice interactions ($J_{AB}$: A-O-B) is much stronger than the intra-sublattice interactions ($J_{AA}$: A-O-A and $J_{BB}$: B-O-B) in spinel ferrites with collinear ferrimagnetic structure. Extensive works on certain spinel ferrites have been carried out for the last few decades, because of their theoretical understanding and potential applications in science and technology. For example, $Fe_3O_4$ (magnetite) derived compounds ($Fe_{3-x}M_xO_4$: M is magnetic or non-magnetic elements-Co, Mn, Zn, etc.) [2-4] have drawn a lot of research interests for their exhibition of many unusual physical properties [5-7], including high magnetic moment, magneto-resistance, half metallic behaviour. In particular, the cobalt substituted magnetite ($Fe_{3-x}Co_xO_4$) could be more attractive due to typical anisotropy character of Co ions. However, most of the reports [4, 8, 9] are limited to the Fe rich regions (x ≤1) of $Fe_{3-x}Co_xO_4$ series. A few reports [10] are only available for the Co rich regions (x ≥ 2), although these compounds have potential applications in chemical sensors [11, 12], catalytic activity [13] and photo-conductive materials [14-16]. $Co_3O_4$ is an antiferromagnet ($T_N$~30 K) but its derivative compound $Co_{3-x}M_xO_4$ [M = Al, Mn, Fe] are generating a lot of research interest in recent times [17, 18]. The study of Co rich compounds could be of interesting in view of the diversity in magnetic properties (ferrimagnet, antiferromagnet and antiferromagnetic spin glass) and magneto-transport phenomena (colossal magnetoresistance) in $Co_xMn_{3-x}O_4$ (0 <x <3) spinel oxides [19, 20].

In the present work, we focus on the spinel compound $Co_2FeO_4$. The experimental work on this compound is very limited. The probable reason could be the spinel phase stability over a small temperature range about 900°C [21, 22]. A few works may be mentioned here, which indicated that the distribution of cations in $Co_2FeO_4$ is very much sensitive on the preparation techniques. Murray *et al.* [23] suggested that the probable distribution of cations in $Co_2FeO_4$ could be $(Co^{2+}_{0.7}Fe^{3+}_{0.3})_A[Co^{2+}_{0.3}Fe^{3+}_{0.7}Co^{3+}]_BO_4$. Based on the Mössbauer study [24], Magnetic measurement [25] and Neutron diffraction study [22], the estimated distribution of cations in $Co_2FeO_4$ are $(Co^{2+}_{0.55}Fe^{3+}_{0.45})_A[Co^{2+}_{0.45}Fe^{3+}_{0.55}Co^{3+}]_BO_4$, $(Co^{2+}_{0.6}Fe^{3+}_{0.4})_A[Co^{2+}_{0.4}Fe^{3+}_{0.6}Co^{3+}]_BO_4$, and $(Co^{2+}_{0.82}Fe^{3+}_{0.18})_A[Co^{2+}_{0.18}Fe^{3+}_{0.82}Co^{3+}]_BO_4$, respectively. The presence of multi valence cations in B sublattice may attribute to a good magnetic semiconducting property in $Co_2FeO_4$, although the experimental results [16-18, 26] indicated an increase of band gap energy (0.14 eV to 1.1 eV) with the increase of Co content in $Fe_{3-x}Co_xO_4$. It is also noted that $Co_2FeO_4$ is prepared in most of the reports using different chemical methods, e.g., coprecipitation [21, 27] and combustion reaction [28]. The fact is that physical properties may depend upon the preparation technique as well as particle morphology (size, shape) of the sample [29].

Considering the theoretical interest [10] and technological applications of $Co_2FeO_4$, detailed information of the correlation between crystal structure and physical properties are very much essential. However, limited information on this aspect of $Co_2FeO_4$ can be obtained due to lack of sufficient experimental work. Here, we emphasize on the understanding of structural phase stabilization and magnetic properties of mechanical alloyed $Co_2FeO_4$ nanoparticles.

## II. EXPERIMENTAL
### A. Sample preparation

Stoichiometric amounts of high purity (>99.5) $Fe_2O_3$ and $Co_3O_4$ were mixed to obtain the required compound. The mixture was ground using mortar and pestle for 2 hours. The mixed powder was mechanical alloyed using Fritsch Planetary Micro Mill "Pulversette 7". The milling was carried out in a 50 ml silicon nitride bowl in atmospheric conditions. The mass ratio of the ball (10 mm Silicon Nitride and 5 mm Tungsten Carbide) and material was maintained to 4:1. The milling was continued at rotational speed 300 rpm up to 100 hours with intermediate stopping for proper mixing and monitoring the phase evolution of the alloyed compound. After 100 hours milling, the alloyed powder was made into pellets of several batches. Each pellet was annealed at different temperatures in the range 700°C to 1000°C. The annealed samples have been denoted as SX, where X =70, 80, 86, 90, 95 and 100 for annealing temperature at 700°C (3 hours), 800°C (6 hours), 860°C (12 hours), 900°C (12 hours), 950°C (12 hours) and 1000°C (3 hours), respectively.

**B. Sample Characterization**

The crystallographic phase of the samples has been examined from the X-ray diffraction (XRD) spectrum using X-Pert PANalytical diffractometer. The spectrum of each sample was recorded at 300 K using Cu K$_\alpha$ radiation in the 2θ range 10 to 90 degrees with step size 0.01 degrees. The scanning electron microscope (SEM) (model: HITACHI S-3400N, Japan) was employed to study the surface morphology of the samples. The elemental analysis of the samples was carried out using Energy Dispersive analysis of X-ray (EDAX) spectrometer (Thermo electron corporation Instrument, USA). Magnetic properties of the samples were investigated from the dc magnetization measurements using Vibrating Sample Magnetometer (Model: 7404 LakeShore, USA), with high temperature oven attachment. The temperature dependence of magnetization was carried out at 1 kOe magnetic field by increasing the temperature from 300 K to 900 K (ZFC mode) and reversing back the temperature to 300 K in the presence of same applied field 1 kOe (FC mode). It should be noted that the ZFC mode denoted here is little bit different from the conventional zero filed cooling (ZFC) measurement, where the sample is first cooled without applying magnetic field from the temperature greater than $T_C$ to the temperature lower than $T_C$ and magnetization measurement starts in the presence of magnetic field by increasing the temperature. The field dependence of magnetization of the samples was measured at 300 K in the field range ± 15 kOe. It may be mentioned that the VSM was calibrated using a standard Ni ferromagnet before starting the measurement of sample.

**III. RESULTS AND DISCUSSION**

The X-ray diffraction pattern of $Co_2FeO_4$ samples are shown in Fig. 1. The XRD spectrum of 100 hours milled sample (data not shown) is largely dominated by spinel phase, along with a small fraction of unreacted α-$Fe_2O_3$ phase. There is no trace of peak lines corresponds to α-$Fe_2O_3$ phase after annealing the alloyed sample at different temperatures. It is found (Fig .1) that XRD peaks of the annealed samples, except 900$^o$C and 950$^o$C, are splitted into two components. We are confirmed, from the comparative peak positions of XRD spectra of our samples and spectrum of standard spinel compounds ($Co_3O_4$, $CoFe_2O_4$) using the same X-Ray diffractometer, that additional peak lines are not contributed due to Cu K$_{\alpha 2}$ radiations. The splitted peaks are, in fact, matching to the spectra of two cubic spinel phases with a shift in 2θ values. The splitting of spectral lines is noted in Fig. 1 (left hand side: 2θ range 10 to 90$^0$) and clearly shown for (311) XRD peak line alone in the right hand side of Fig. 1. The peak positions at higher and lower 2θ values are denoted at $2θ_2$ and $2θ_1$, respectively. As the annealing temperature increases from 700°C to 1000°C, the peaks at $2θ_1$ and $2θ_2$ are coming closer to each other and there is no splitting at 900°C (S90) sample, suggesting single phased compound. Although there is no clear splitting in (311) XRD line for 950°C (S95) sample (*i.e., the sample is appeared to be in single phase*), but there is a tendency of minor splitting in other XRD lines of 950°C sample in comparison with 900°C sample. This indicates that a minor secondary phase may coexist at 950°C. The peaks correspond to $2θ_1$ and $2θ_2$ are again well separated at 1000°C, indicating the reappearance of phase separation in the material. The results of our mechanical alloyed and subsequent annealed samples also confirm the earlier reports [21, 22] that $Co_2FeO_4$ is stabilized into a single cubic spinel phase at about 900°C (S90), although our preparation technique (mechanical alloying) is completely different from the reported works. It may be mentioned that we did not find any significant change in the XRD spectrum when slow cooled sample is compared with the direct air quenched samples. Smith *et al.* [24] has reported the same effect from X-ray diffraction and Mössbauer measurements.

The lattice parameter was calculated by matching the XRD peaks at $2θ_1$ and $2θ_2$ separately with cubic spinel structure, considering the coexistence of two spinel phases in the spectrum. The lattice parameter corresponds to $2θ_1$ and $2θ_2$ are denoted as $a_1$ and $a_2$, respectively. The data are shown in Fig. 2a. The lattice parameter $a_1$ decreases as the annealing temperature increases from 700 to 900$^0$C, unlike the increase of $a_2$. Both ($a_1$ and $a_2$) are coinciding for 900°C (S90) sample. In the absence of clear splitting for the 950°C (S95) sample, we have fitted the spectrum assuming the single phase. The lattice parameter of 950°C sample is little bit higher in comparison with 900°C sample. The lattice parameters $a_1$ and $a_2$, again, separated for the 1000°C sample with $a_1$ higher than $a_2$. The calculated lattice parameter $a$ (~8.24 Å) for the 900°C sample is in good agreement with the reported data both from theoretical calculation ($a$ =8.27 Å) [10] and experimental work ($a$ =8.24 Å) [24]. The lattice parameter for single phased $Co_3O_4$ and $CoFe_2O_4$ is found ~ 8.082 Å and 8.40 Å, respectively. On the other hand, $a_1$ and $a_2$ are ~

8.37 Å and 8.13 Å for S70 sample and ~ 8.31 Å and 8.16 Å for S100 sample. In between, the system stabilizes to single phased cubic spinel structure of $Co_2FeO_4$ with ***a*** ~8.24 Å. Viewing the different values of ***$a_1$*** and ***$a_2$***, we suggest that ***$a_1$*** (corresponds to peaks at $2\theta_1$) is contributed from Fe-rich cubic spinel structure and ***$a_2$*** (corresponds to peaks at $2\theta_2$) is contributed from Co-rich cubic spinel structure. We have attempted to estimate the fraction of coexisting two phases from the relative peak heights at $2\theta_1$ and $2\theta_2$ positions, assuming that the addition of two peak heights will contribute to the total peak height of single phase cubic spinel structure. Fig.2b shows the % height of splitted peaks at $2\theta_1$ and $2\theta_2$ with respect to (311), (004) and (333) peaks of single phased cubic spinel structure (S90 sample). We observed that Fe-rich phase increases from ~60% (S70) to 100% (S90 and S95) by decreasing the Co-rich phase from ~40% to 0%. For S100 sample, the Co-rich phase increases by decreasing the Fe-rich phase. The change of relative peak heights of the two phases with annealing temperature is found to be same within the error limit of peak height determination for all the three peaks (Fig. 2b).

The particle size of the samples was determined using Debye-Scherrer formula: $\langle d \rangle = 0.89\lambda/\beta\cos\theta$ ($\lambda$ the wavelength of the X-ray, $2\theta$ corresponds to position of peak height, $\langle d \rangle$ is particle size, $\beta$ is the full width at half maximum of peak height) on four ((311), (004), (333), (044)) XRD peaks at $2\theta_1$. The particle size (in Table I) showed the usual increase (~25 nm to 45 nm) with annealing temperatures, as an effect of thermal induced grain growth kinetics. The SEM pictures of selected samples are shown in Fig 3 (a-d). The general observation from the Fig. 3 (a-d) is that thermal annealing increases the particle size of the samples. The particles are in agglomerated state in the as alloyed sample (MA100), whereas a better homogeneity in the particle size distribution is observed in the annealed samples. The elemental composition of the samples is determined from the EDAX spectrum over a selected zone. The spectrum of each sample was recorded for 10 points and the selected spectrum is shown in Fig. 3e-h. The expected atomic ratio of Co and Fe in $Co_2FeO_4$ must be 2:1. The spectrum of MA100 sample (Fig. 3e) suggested an inhomogeneous elemental distribution. This means some points are having more Fe atom (Co: Fe=2:1.35) and other points are having less Fe atoms (Co: Fe= 2:0.85) compared to the expected value (2:1). The Fe rich region in MA100 is probably due to the presence of a fraction of unreacted $Fe_2O_3$ in the sample, as seen from the XRD spectrum. The chemical inhomogeneity is again noted in the EDAX spectrum of annealed samples, except S90 sample (Fig. 3g). For S80 sample (Fig. 3f), there are some points which are Co-rich (Co: Fe~2.078:1) and some points are Co-deficient (Co:Fe~1.78:1). The S100 sample (Fig. 3h) also showed similar atomic distribution with Co rich (Co: Fe ~3.053:1) zone and Co-deficient (Co:Fe ~1.64:1) zone. On the other hand, the Co and Fe atoms are almost homogeneously distributed over the selected zone of S90 sample and the atomic ratio (Co:Fe~1.95:1) is close to the expected value (2:1). The atomic percentage of the non-magnetic impurity atoms (Si ~ 0.7 % and W ~ 0.4 %) from the milling bowl and balls is very insignificant in the alloyed as well as in the annealed samples.

The temperature (T) dependence of magnetization, measured under ZFC and FC modes at 1 kOe, is shown in Fig. 4a-e. The magnetic irreversibility between FC magnetization (MFC) and ZFC magnetization (MZFC) is seen for all the samples. The nature of irreversibility is distinct for individual sample. The mixed ferrimagnetic phase of the samples, except S90, is reflected from the magnetization plots. The paramagnetic to ferrimagnetic ordering temperature ($T_{C1}$) of one phase is determined from the inflection point of MZFC curves. It may be noted that the irreversibility between MFC and MZFC occured at temperature that is much higher than $T_{C1}$. We define the irreversibility point as the paramagnetic to ferrimagnetic transition temperature ($T_{C2}$) of the second ferrimagnetic phase ($T_{C2} > T_{C1}$). The coexistence of two magnetic phases is also indicated in the MFC(T) curve. The MFC curves continued to increase with decreasing the temperature below irreversibility point, but a change of slope is marked near to $T_{C1}$ of the sample. The signature of changing slope was also verified from the first order derivative of MFC(T) (data are not shown). The magnetic irreversibility occurs at T ≤ 450 K for S90 sample and there is no change of slope in MFC(T) curve of the sample. This indicates a remarkable change in the ferrimagnetic behaviour of S90 sample in comparison with other samples. The $T_{C1}$ of S90 sample, determined from the inflection point of MZFC(T), is ~ 453 K. This is consistent with the reported value ~ 450 K of single phase $Co_2FeO_4$ [22]. The typical MFC and MZFC behaviour of the samples, associated with mixed magnetic phases, are also understood from the temperature dependence of normalized thermoremanent magnetization (NTRM = $\Delta M(T)/ \Delta M(300 K)$, where $\Delta M$ = MFC-MZFC) data (Fig. 4f). We, now, look at the magnetic behaviour of the samples below $T_{C1}$. The zero field cooled magnetization exhibited blocking behaviour below the temperature $T_m$ (~ 420 K for S80, ~ 360 K for S86 and ~ 400 K for S100 samples, respectively). The continuous increase of MZFC down to 300 K for S90 and S95 samples without blocking of magnetization suggests that the possible blocking behaviour for these two nanoparticle samples might occur below room temperature.

We have seen some interesting magnetic features of the samples from the field (H) dependence of magnetization (M) curve at 300 K (Fig. 5a). The magnetization of the samples, after rapid increase within 5 kOe, is tending to saturate at higher magnetic field. The feature suggests a typical long ranged ferrimagnetic character of the samples irrespective of phase stability. The spontaneous magnetization ($M_S$) at 300 K is calculated from the extrapolation of high field magnetization data to H = 0 value. The $M_S$ value (Table I) shows decreasing trend with increasing the annealing temperature from $800^0$C (~23.5 emu/g) to $900^0$C (~16 emu/g). After attaining the minimum value for S90 sample, the $M_S$ value again is increasing with annealing temperature. The variation of $M_S$ with annealing temperature shows a close relation to the variation of lattice parameter $a_1$ (due to Fe rich spinel phase). We, further, try to understand such correlation from the following arguments. The calculated magnetic moment 16 emu/g (~0.68 $\mu_B$ per formula unit of $Co_2FeO_4$) for S90 sample at 300 K is comparable to the reported value (0.70 $\mu_B$) for single phase compound [22]. We also noted that magnetic moment of S80 and S100 samples ~1.0 $\mu_B$ per formula unit of $Co_2FeO_4$ is well below of the magnetic moment ~4.2 $\mu_B$ per formula unit of $CoFe_2O_4$. On the other hand, higher Curie temperature ($T_{C2}$) of the bi-phase samples, e.g., ~752 K for S80 and ~772 K for S100, are close to the Curie temperature ($T_C$ ~785 K) of single phase $CoFe_2O_4$ [30]. This means the ferrimagnetic phase with Curie temperature $T_{C2}$ is associated with the phase that may not be due to typical $CoFe_2O_4$, but definitely due to a Fe-rich spinel phase coexisting with Co-rich (low magnetic moment) phase. Now, we examine the M-H loop of the samples in Fig. 5 (b-f). Although measurement is done within H = ± 15 kOe, the data are shown within H = ± 7 kOe for clarity of the Figure. It is interesting to note that M-H loop for S90 and S95 samples is distinguished with more symmetric nature in comparison with other annealed samples. Such interesting magnetic feature clearly reflects the undergoing competitive spin order of two ferrimagnetic domains, arising from the coexistence of two different type cubic spinel phases in the material. The calculated values of coercivity ($H_C$) and remanent magnetization ($M_R$) from M-H loop (also known as hysteresis loop) of the samples are shown in Table I. The coercivity ($H_C$) and remanent magnetization ($M_R$) are showing similar kind of variation with annealing temperature similar to $M_S$, except $H_C$ and $M_R$ attains minimum value for S95 and S90 samples, respectively. The small increase of $M_R$ in S86 sample compared to S80 sample is most probably related to the more asymmetric shape of the M-H loop in S86 sample. The $M_R/M_S$ ratio (~0.42, 0.55, 0.46, 0.40 and 0.44 for S80, S86, S90, S95 and S100 samples, respectively) suggests that about 40 to 50% of the spontaneous magnetization is retained in the material as soon as the maximum applied field (+15 kOe) is reduced to zero. The variation of dM/dH with applied magnetic field (Fig. 6) showed peaks, which are almost symmetric about the H = 0 axis. The peaks are separated by magnetic field $2H_m$, as shown for sample S80. It may be noted (Table I) that the peak position at $H_m$, i.e., the inflection point in M-H curve, is very close to the coercivity ($H_C$) of the samples, except S86 sample that has shown more asymmetric M-H loop. The peak height of dM/dH at $H_m$ (~ 0.0165, 0.021, 0.026, 0.060 and 0.027 in emu/g Oe unit for S80, S86, S90, S95 and S100 samples, respectively) is increasing to attain the maximum value for S95 sample in comparison with other samples. We noted that the initial susceptibility $(dM/dH)_{H \to 0}$ of the sample is nearly half of the dM/dH peak height at $H_m$, i.e, at the inflection point of the M-H curves. The peaks are comparatively narrowed for S90 and S95 samples. This, further, indicates that ferrimagnetic domains are attaining a better homogeneous structure in the temperature range 900-950$^0$C and can be realized from the single phase character of the XRD spectrum.

## IV. SUMMARY AND CONCLUSIONS

The present work successfully applied the novel technique of mechanical alloying to synthesize $Co_2FeO_4$ spinel oxide. The structural phase evolution of the synthesized compound with annealing temperature consistent with earlier reports, based on chemical routed samples. The experimental results suggest that crystal structure of the samples, except 900$^0$C annealed sample, is separated into the structure of two cubic spinel phase, consisting of Co-rich and Fe-rich phase. The blocking of MZFC below $T_m$ is related to particle size of the samples in nanometer range. However, the material does not exhibit the conventional variation (either increase or decrease) of $T_m$ with the increase of particle size, rather the magnetic blocking of nanoparticles is affected by the phase stability of cubic spinel structure. The detailed low temperature (below 300 K) study for understanding the magnetic blocking behaviour is not within the scope of the present work. The high temperature study indicated that the samples are ferrimagnet, irrespective of structural phase stability. The co-existence of two ferrimagnetic phases in the samples, except $900^0C$, further supported the structural phase separation and confirmed a strong correlation between cubic spinel structure and magnetism in $Co_2FeO_4$ spinel oxide. The understanding of such correlation may be applied to the other oxide materials, including Perovskites, which have shown phase separation phenomena and associated magneto-transport properties.

In conclusion, the physical picture that might occur in the system is that Co atoms are not uniformly diffusing to the Fe lattice positions at the temperatures differing from $900^0C$, resulting in the separation of Co-rich and Fe- rich cubic spinel structures in the same material. As the temperature approaches to $900^0C$, the Co-rich phase is melting into the Fe-rich phase to form a homogeneous solid solution of $Co_2FeO_4$ spinel oxide. The ferrimagnetic properties are strongly correlated to the structural phase instability of the material.

**Acknowledgment:** The authors thank to CIF, Pondicherry University for providing experimental facilities.

**Table. I**. Curie temperatures ($T_{C1}(K)$ and $T_{C2}(K)$) and blocking temperature ($T_m(K)$), Spontaneous magnetization ($M_S$), Coercivity ($H_C$), remanent magnetization ($M_R$) and critical field ($H_m$) of $Co_2FeO_4$ samples.

| Sample | Particle Size (nm) | $T_{C1}$ (K) | $T_{C2}$ (K) | $T_m$ (K) | $M_S$ (emu/g) | $H_C$ (Oe) | $M_R$ (emu/g) | $H_m$ (Oe) |
|---|---|---|---|---|---|---|---|---|
| S80 | 25 | 645 | 752 | 423 | 23.5 | 696 | 9.85 | 700 |
| S86 | 29 | 555 | 750 | 358 | 19 | 540 | 10.53 | 415 |
| S90 | 31 | 453 | -- | -- | 16 | 280 | 7.28 | 300 |
| S95 | 36 | 449 | 560 | -- | 21.6 | 125 | 8.57 | 130 |
| S100 | 42 | 581 | 772 | 396 | 22 | 430 | 9.75 | 460 |

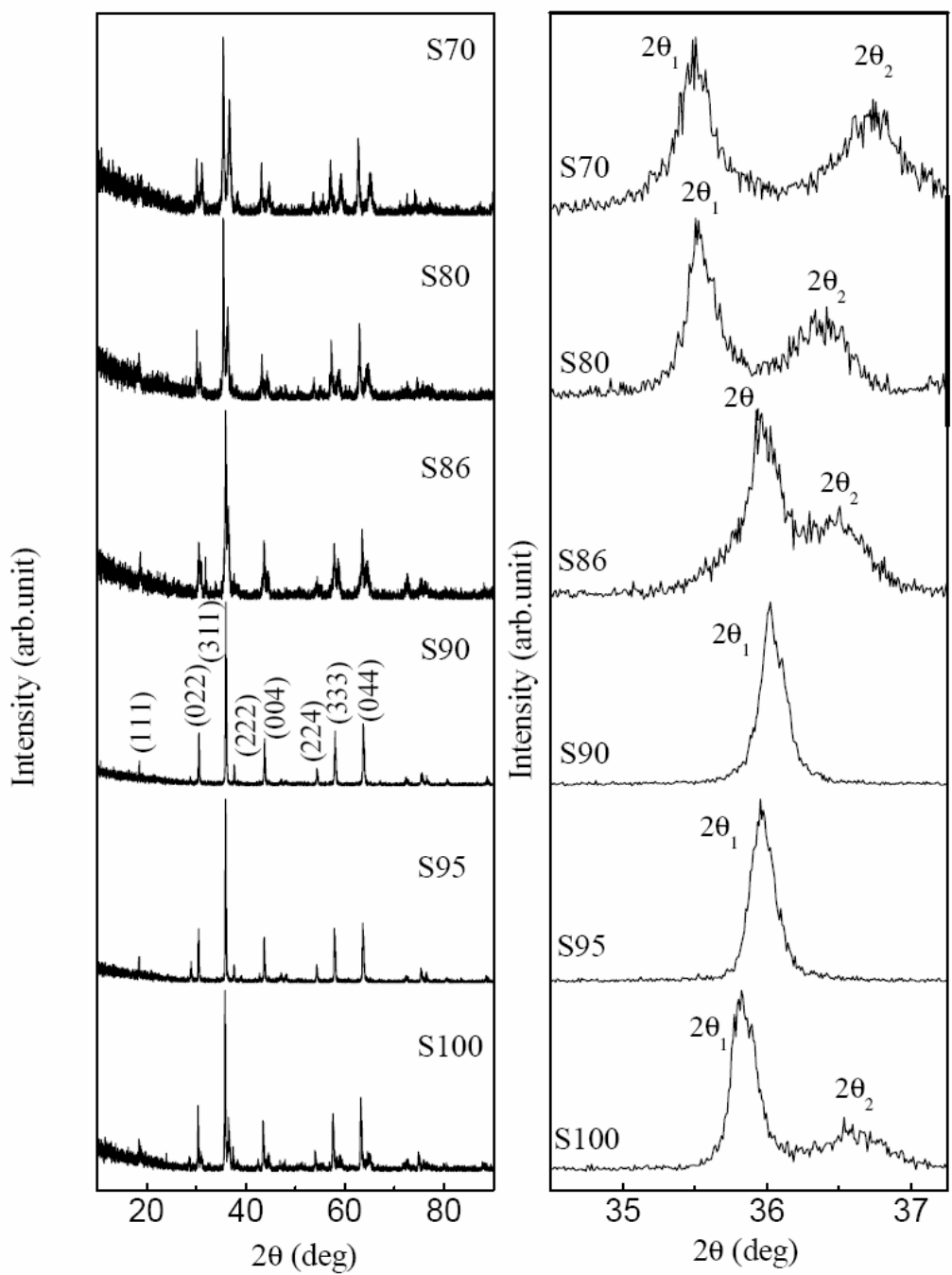

Fig. 1. XRD spectra of the annealed sample in the range 10-90 degree (left hand side) and in the range 34-37.5 degree (right hand side) to indicate the splitting of 311 peak at $2\theta_1$ and $2\theta_2$ positions.

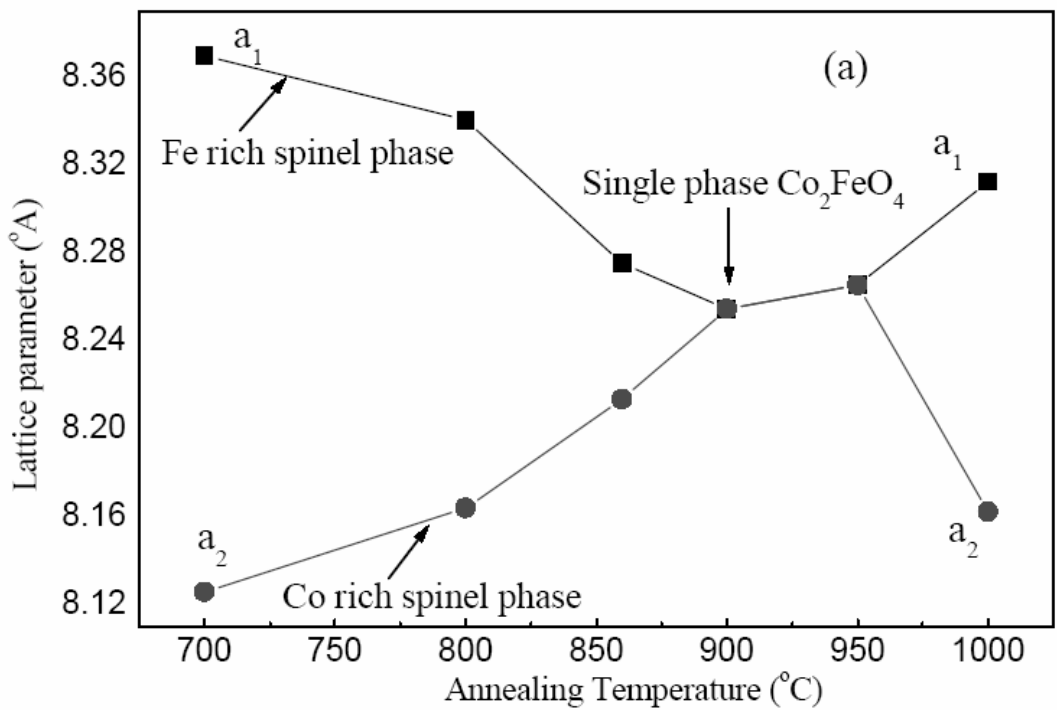
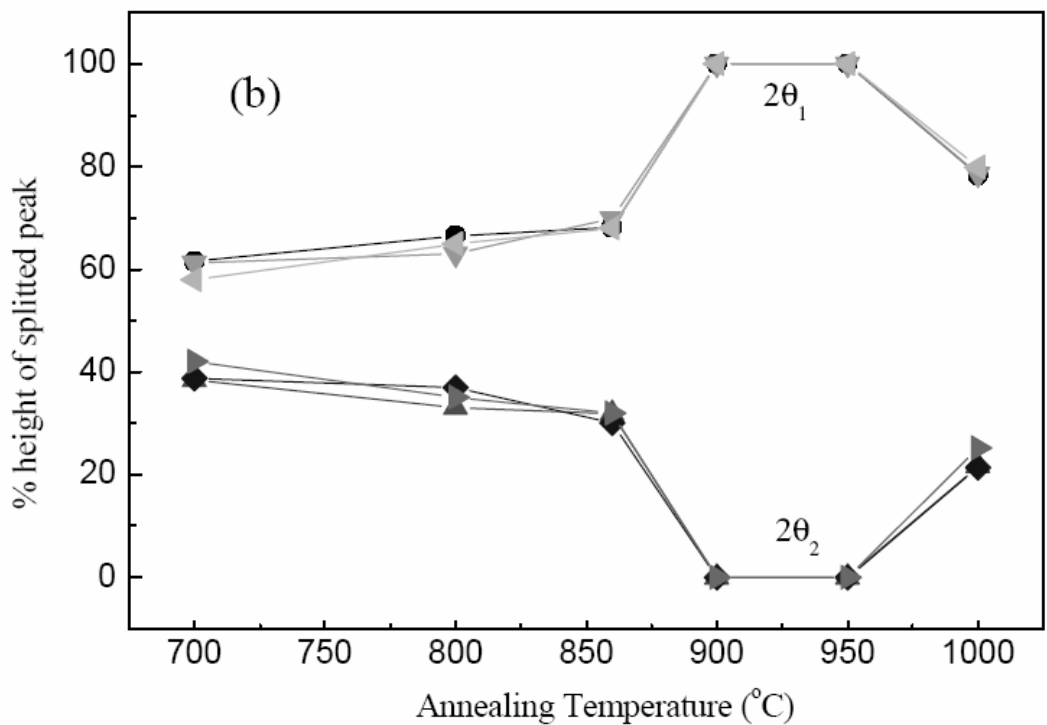

Fig. 2. (Colour online) Lattice parameters $a_1$ and $a_2$ and % of (311) peak height at $2\theta_1$ and $2\theta_2$ positions of annealed samples are shown in (a) and (b), respectively.

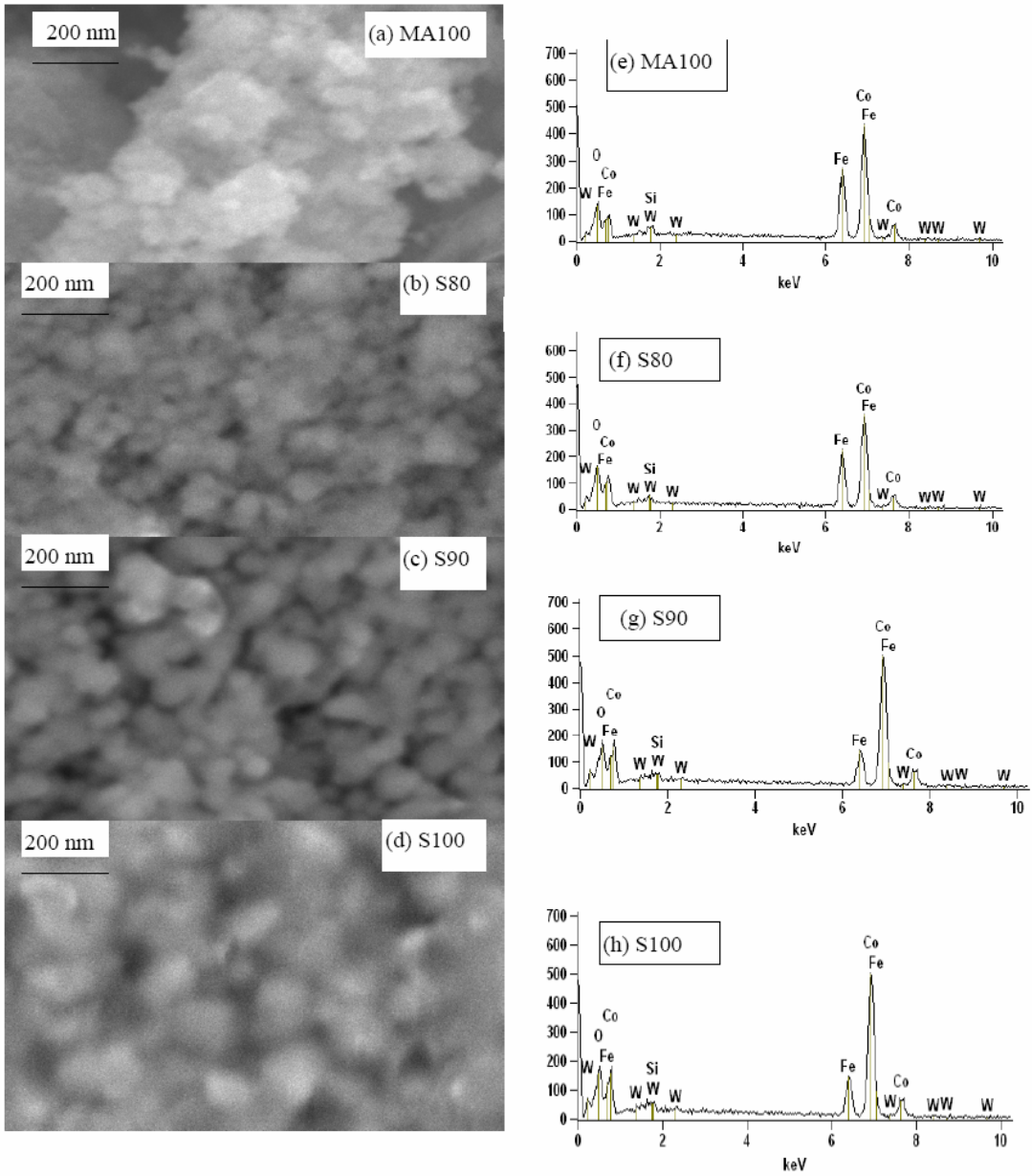

Fig. 3. SEM pictures (in a-d) of MA100, S80, S90 and S100 samples and corresponding EDAX spectrum (in e-h).

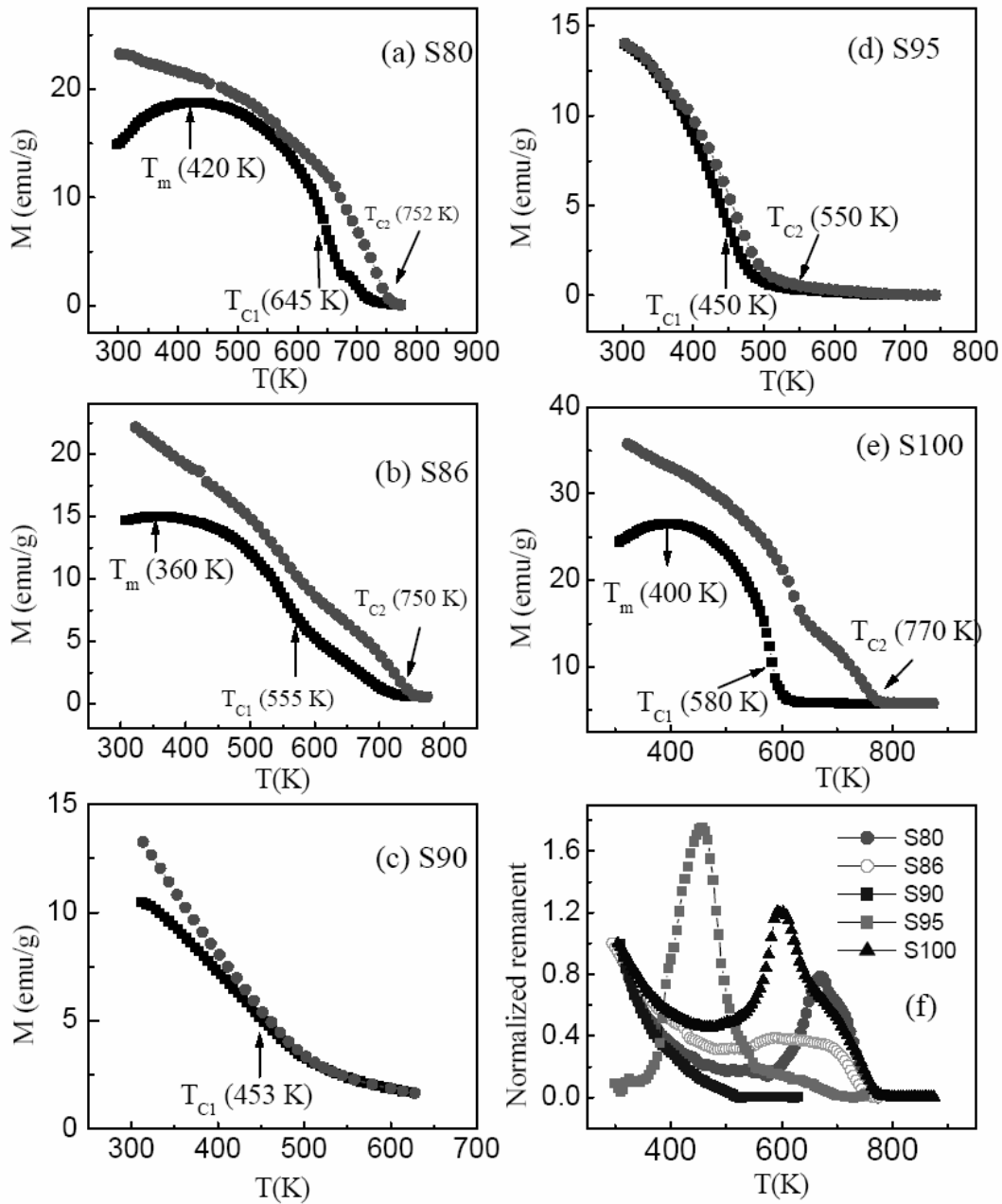

Fig. 4. (Colour online) MZFC (closed symbol) and MFC (open symbol) vs. T for S80, S86, S90, S95 and S100 samples (in a-e), measured at 1 kOe. $T_{C1}$, $T_{C2}$ and $T_m$ are defined in text. The normalized thermoremanent magnetization (NTRM = $\Delta M(T)/\Delta M(300\ K)$, where $\Delta M$ = MFC-MZFC) of the samples are shown in Fig. 4f. TRM is not normalized for S95 sample to maintain the proper scale.

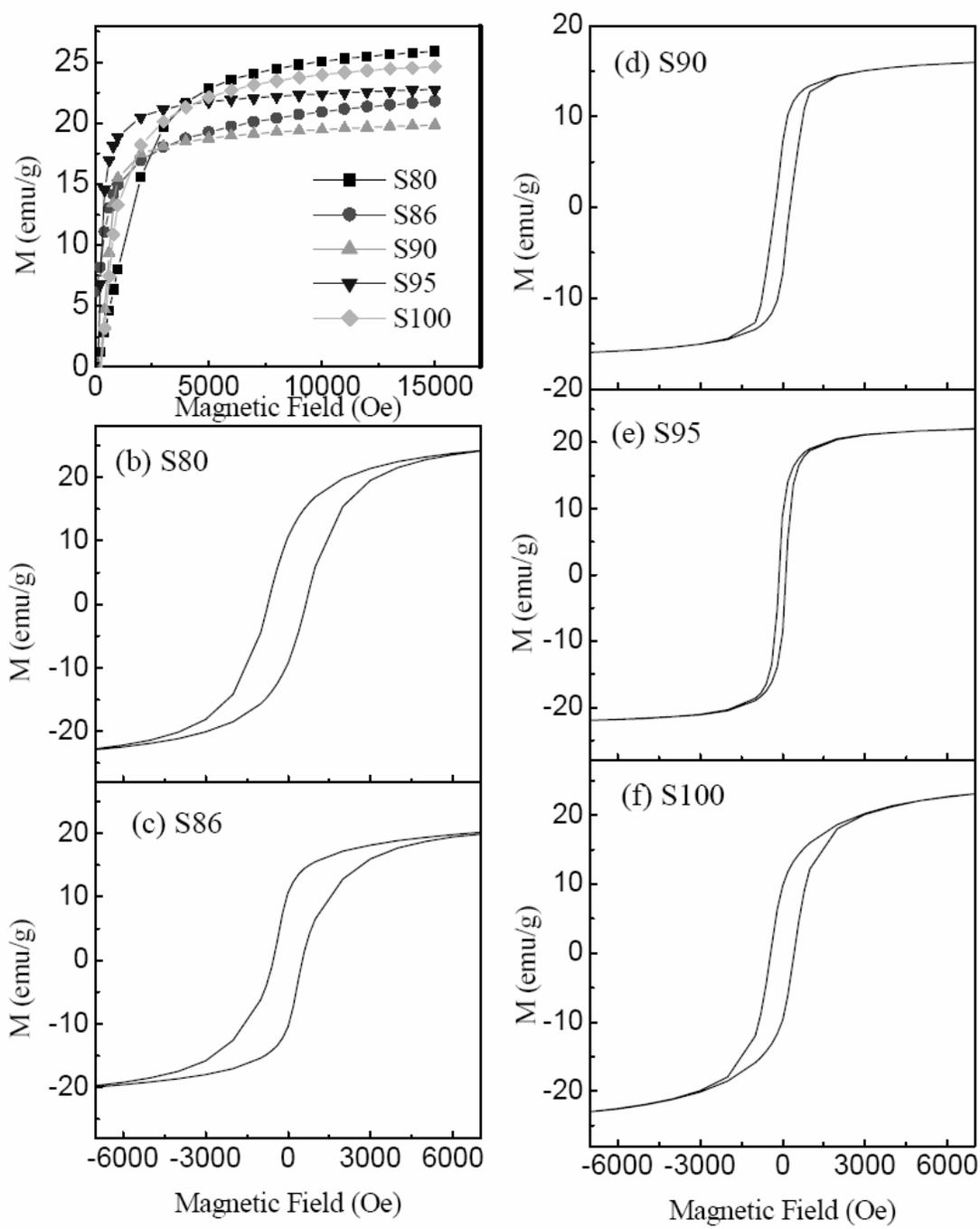

Fig. 5. (Colour online) Field dependence of magnetization at 300 K (in a). The M-H loops of the samples are shown in b-f.

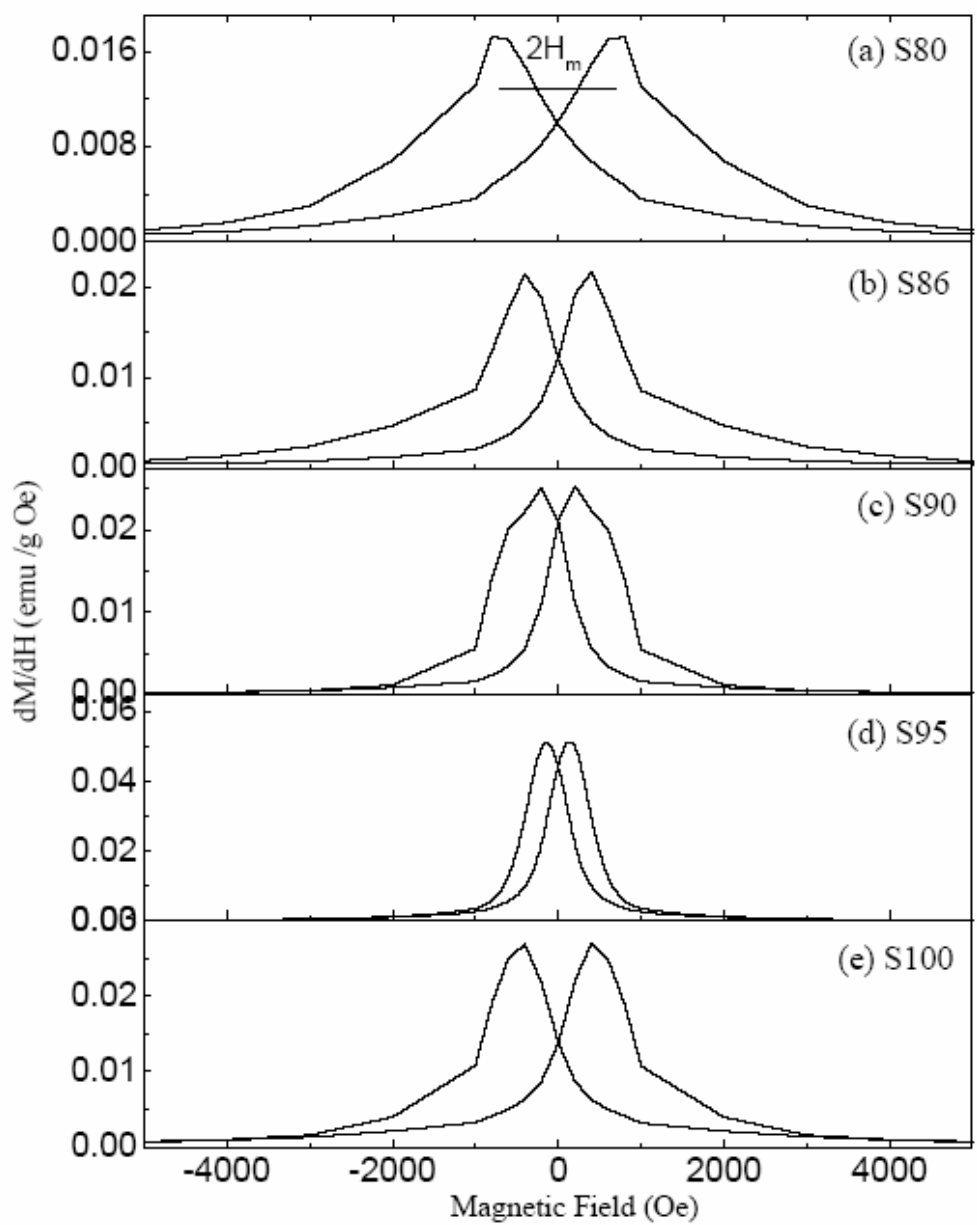

Fig. 6. Field dependence of dM/dH of the samples at 300 K. The separation of the peaks is represented by the field $2H_m$.